\documentclass[11pt]{article}
\usepackage{amsmath, amsthm, amssymb, color, graphicx, psfrag, tabularx, hyperref, bbm, nicefrac}

\newcommand{\be}{\begin{equation}}
\newcommand{\ee}{\end{equation}}
\newcommand{\bea}{\begin{eqnarray}}
\newcommand{\eea}{\end{eqnarray}}
\newcommand{\beas}{\begin{eqnarray*}}
\newcommand{\eeas}{\end{eqnarray*}}

\newcommand{\R}{\ensuremath{\mathbb{R}}}
\newcommand{\N}{\ensuremath{\mathbb{N}}}
\newcommand{\F}{\ensuremath{\mathbb{F}}}
\newcommand{\IH}{\ensuremath{\mathbb{H}}}

\newcommand{\crl}[1]{\ensuremath{ \left\{ #1 \right\} }}
\newcommand{\edg}[1]{\ensuremath{ \left[ #1 \right] }}
\newcommand{\brak}[1]{\ensuremath{\left( #1 \right)}}

\newcommand{\p}{\ensuremath{\mathbb{P}}}

\newcommand{\E}{\ensuremath{\mathbb{E}}}

\newtheorem{theorem}{Theorem}
\newtheorem{remark}[theorem]{Remark}

\title{Pricing and hedging American-style options with deep learning}
\author{Sebastian Becker\footnote{RiskLab, ETH Zurich, Switzerland}
\and Patrick Cheridito$^*$
\and Arnulf Jentzen\footnote{Faculty of Mathematics and Computer Science, 
University of M\"unster, Germany}
\footnote{A. Jentzen acknowledges support from the DFG through Germany's Excellence Strategy EXC 2044-390685587, 
Mathematics M\"unster: Dynamics - Geometry - Structure.}}

\date{}

\begin{document}
\maketitle

\begin{abstract}
In this paper we introduce a deep learning method for pricing and hedging
American-style options. It first computes a candidate optimal stopping policy.
From there it derives a lower bound for the price. Then it calculates 
an upper bound, a point estimate and confidence intervals. Finally, it constructs an approximate 
dynamic hedging strategy. We test the approach on different specifications of a
Bermudan max-call option. In all cases it produces highly accurate prices and dynamic hedging 
strategies with small replication errors.\\[2mm]
\end{abstract} 

\section{Introduction}
\label{sec:intro}

Early exercise options are notoriously difficult to value. For up to three 
underlying risk factors, tree based and classical PDE approximation methods usually yield good numerical 
results; see, e.g., \cite{Hull,FV02, RW12} and the references therein. To treat higher-dimensional problems, various 
simulation based methods have been developed; see, e.g., \cite{Till93, BM95, Ca96, A00, LS01, TVR01, G03, BG04, BPP05,
KS06, EKT07, BC08, BS08, JO15}. \cite{HK04, KKT10} have already used shallow\footnote{meaning
feedforward networks with a single hidden layer} neural networks to estimate continuation values. More recently, in
\cite{SS18} optimal stopping problems in continuous time have been solved
by approximating the solutions of the corresponding free boundary PDEs with deep neural networks. 
In \cite{DOS, BCJW} deep learning has been used to directly learn optimal stopping strategies.
The main focus of these papers is to derive optimal stopping rules and accurate price estimates.

The goal of this article is to develop a deep learning method which learns
the optimal exercise behavior, prices and hedging strategies from samples 
of the underlying risk factors. It first learns a candidate optimal stopping strategy by regressing 
continuation values on multilayer neural networks. Employing the learned stopping strategy on 
a new set of Monte Carlo samples gives a low-biased estimate of the price.
Moreover, the candidate optimal stopping strategy can be used to construct an approximate solution 
to the dual martingale problem introduced by \cite{R02} and \cite{HK04}, yielding 
a high-biased estimate and confidence intervals for the price. In the last step, our method
learns a dynamic hedging strategy in the spirit of \cite{HJE} and \cite{BGTW19}. But here, the continuation value
approximations learned during the construction of the optimal stopping strategy can be used
to break the hedging problem down into a sequence of smaller problems that learn the hedging 
portfolio only from one possible exercise date to the next. Alternative ways of computing 
hedging strategies consist in calculating sensitivities of option prices (see e.g., \cite{BPP05, BW12, JO15}) 
or approximating a solution to the dual martingale problem (see e.g., \cite{R02, R10}).

Our work is related to the preprints \cite{LL19} and \cite{CW19}. \cite{LL19} also uses neural network regression 
to estimate continuation values. But the networks are slightly different. While \cite{LL19} 
works with leaky ReLU activation functions, we use $\tanh$ activation. Moreover, \cite{LL19} 
studies the convergence of the pricing algorithm as the number of simulations and 
the size of the network go to infinity, whereas we calculate a posteriori guarantees for the 
prices and use the estimated continuation value functions to implement efficient 
hedging strategies. \cite{CW19} proposes an alternative way of calculating 
prices and hedging strategies for American-style options by solving BSDEs. 

The rest of the paper is organized as follows. In Section \ref{sec:st} we describe 
our neural network version of the Longstaff--Schwartz algorithm to 
estimate continuation values and construct a candidate optimal stopping 
strategy. In Section \ref{sec:pricing} the latter is used to derive lower and upper bounds
as well as confidence intervals for the price. Section \ref{sec:hedging} 
discusses two different ways of computing dynamic hedging strategies. 
In Section \ref{sec:ex} the results of the paper are applied to price and hedge 
a Bermudan call option on the maximum of different underlying assets. 
Section \ref{sec:conclusion} concludes.

\section{Calculating a candidate optimal stopping strategy}
\label{sec:st}

We consider an American-style option that can be exercised at any one of finitely\footnote{This covers 
Bermudan options as well as American options that can only be exercised at a given time each day. 
Continuously exercisable options must be approximated by discretizing time.} many times $0 = t_0 < t_1< \dots < t_N = T$. 
If exercised at time $t_n$, it yields a discounted payoff given by a square-integrable random variable 
$G_n$ defined on a filtered probability space $(\Omega, {\cal F}, \mathbb{F} = ({\cal F}_n)_{n=0}^N, \p)$. We assume that 
${\cal F}_n$ describes the information available at time $t_n$ and $G_n$ is of the form 
$g(n,X_n)$ for a measurable function $g \colon \crl{0, 1, \dots, N} \times \R^d \to [0, \infty)$ and a 
$d$-dimensional $\mathbb{F}$-Markov process\footnote{That is,  $X_n$ is ${\cal F}_n$-measurable, and
$\E [f(X_{n+1}) \mid {\cal F}_n] = \E [f(X_{n+1}) \mid X_n] $
for all $n \le N-1$ and every measurable function $f \colon \R^d \to \R$ such that 
$f(X_{n+1})$ is integrable.} $(X_n)_{n=0}^N$. We assume $X_0$ to be deterministic 
and $\p$ to be the pricing measure. So that the value of the option at time $0$ is given by
\[
V = \sup_{\tau \in {\cal T}} \mathbb{E} \, G_{\tau},
\]
where ${\cal T}$ is the set of all $\F$-stopping times $\tau \colon \Omega \to \crl{0,1,\dots, N}$.
If the option has not been exercised before time $t_n$, its discounted value at that time is
\be \label{Hn}
V_{t_n} = \mbox{ess\,sup}_{\tau \in {\cal T}_n} \mathbb{E}[G_{\tau} \mid {\cal F}_n],
\ee
where ${\cal T}_n$ is the set of all $\F$-stopping times satisfying $n \le \tau \le N$.

Obviously, $\tau_N \equiv N$ is optimal for $V_T = G_N$. From there, one can recursively construct 
the stopping times 
\be \label{taun}
\tau_n := \begin{cases}
n & \mbox{ if } G_n \ge \mathbb{E} [G_{\tau_{n+1}} \mid X_n]\\
\tau_{n+1} & \mbox{ if } G_n < \mathbb{E} [G_{\tau_{n+1}} \mid X_n].
\end{cases}
\ee
Clearly, $\tau_n$ belongs to ${\cal T}_n$, and it can be checked inductively that
\[
V_{t_n} = \E[G_{\tau_n} \mid {\cal F}_n] = G_n \vee \E[V_{t_{n+1}} \mid X_n] \quad
\mbox{for all } n \le N-1.
\]
In particular, $\tau_n$ is an optimizer of \eqref{Hn}.

Recursion \eqref{taun} is the theoretical basis of the Longstaff--Schwartz method \cite{LS01}.
Its main computational challenge is the approximation of 
the conditional expectations $\mathbb{E} [G_{\tau_{n+1}} \mid X_n]$. It is well known that 
$\mathbb{E} [G_{\tau_{n+1}} \mid X_n]$ is of the form $c(X_n)$, where $c \colon \R^d \to \R$ minimizes the 
mean squared distance $\E \edg{\crl{G_{\tau_{n+1}} - c(X_n)}^2}$ over all 
Borel measurable functions from $\R^d$ to $\R$; see, e.g., \cite{BH85}. The Longstaff--Schwartz 
algorithm approximates $\mathbb{E} [G_{\tau_{n+1}} \mid X_n]$ by projecting $G_{\tau_{n+1}}$ 
on the linear span of finitely many basis functions. But it is also possible to project on a 
different subset. If the subset is given by $c^{\theta}(X_n)$ for a function family
$c^{\theta} \colon \R^d \to \R$ parametrized by $\theta$,
one can apply the following variant\footnote{The main difference between 
this algorithm and the one of Longstaff and Schwartz \cite{LS01} is the 
use of neural networks instead of linear combinations of basis functions. In addition, the sum in \eqref{ce} 
is over all simulated paths, whereas in \cite{LS01}, only in-the-money paths are considered to save computational effort.
While it is enough to use in-the-money paths to determine a candidate optimal stopping rule, we need accurate
approximate continuation values for all $x \in \R^d$ to construct good hedging strategies in Section \ref{sec:hedging}.} 
of the Longstaff--Schwartz algorithm:

\begin{enumerate}
\item[{\rm (i)}]
Simulate\footnote{As usual, we simulate the paths $(x^k_n)$, $k = 1, \dots, K$, independently of each other.} 
paths $(x^k_n)_{n=0}^N$, $k = 1, \dots, K$, of the underlying 
process $(X_n)_{n=0}^N$.
\item[{\rm (ii)}] Set $s^k_N \equiv N$ for all $k$. 
\item[{\rm (iii)}] For $1 \le n \le N-1$, approximate $\mathbb{E} \edg{G_{\tau_{n+1}} \mid X_n}$ with 
$c^{\theta_n}(X_n)$ by minimizing the sum 
\be \label{ce}
\sum_{k =1}^{K} \brak{g(s^k_{n+1}, x^k_{s^k_{n+1}}) - c^{\theta}(x^k_n)}^2
\quad \mbox{over } \theta.
\ee
\item[{\rm (iv)}]
Set 
\[
s^k_n := 
\begin{cases} 
n & \mbox{ if } g(n,x^k_n) \ge c^{\theta_n}(x^k_n)\\
s^k_{n+1} & \mbox{ otherwise.}
\end{cases}
\]
\item[{\rm (v)}]
Define $\theta_0 := \frac{1}{K} \sum_{k =1}^{K} g(s^k_1, x^k_{s^k_1})$,
and set $c^{\theta_0}$ constantly equal to $\theta_0$.
\end{enumerate} 

In this paper we specify $c^{\theta}$ as a feedforward neural network, which in general,
is of the form 
\be \label{nn} 
a^{\theta}_I \circ \varphi_{q_{I-1}} \circ a^{\theta}_{I-1} \circ \dots \circ \varphi_{q_1} \circ a^{\theta}_1,
\ee
where
\begin{itemize}
\item 
$I \ge 1$ denotes the depth and $q_0, q_1, \dots, q_I$ the numbers of nodes in the different layers
\item 
$a^{\theta}_1 \colon \mathbb{R}^{q_0} \to \mathbb{R}^{q_1}, \dots, 
a^{\theta}_I \colon \mathbb{R}^{q_{I-1}} \to \mathbb{R}^{q_I}$
are affine functions,
\item 
For $j \in \mathbb{N}$, $\varphi_j \colon \mathbb{R}^j \to \mathbb{R}^j$ is of the form
$\varphi_j(x_1, \dots, x_j) = (\varphi(x_1), \dots, \varphi(x_j))$
for a given activation function $\varphi \colon \R \to \R$.
\end{itemize}

The components of the parameter $\theta$ consist of the entries of the matrices 
$A_1, \dots, A_I$ and vectors $b_1, \dots, b_I$ appearing in the representation
of the affine functions $a^{\theta}_i x= A_i x + b_i$, $i = 1, \dots, I$. So, 
$\theta$ lives in $\R^q$ for $q = \sum_{i=1}^I q_i(q_{i-1} + 1)$.
To minimize \eqref{ce} we choose a network with $q_I = 1$ and employ a stochastic gradient descent method.

\section{Pricing}
\label{sec:pricing}

\subsection{Lower bound}

Once $\theta_0, \theta_1, \dots, \theta_{N-1}$ have been determined, we set 
$\Theta = (\theta_0, \dots, \theta_{N-1})$ and define
\[
\tau^{\Theta} := \min \crl{n \in \crl{0,1,\dots, N-1}: g(n,X_n) \ge c^{\theta_n}(X_n) }, 
\quad \mbox{where } \min \emptyset \mbox{ is understood as } N.
\]
This defines a valid $\mathbb{F}$-stopping time. Therefore, $L = \E \, g(\tau^{\Theta}, X_{\tau^{\theta}})$ is a lower 
bound for the optimal value $V$. But typically, it is not possible to calculate the expectation exactly. Therefore, we 
generate simulations $g^k$ of $g(\tau^{\Theta}, X_{\tau^{\Theta}})$ based on independent sample 
paths\footnote{generated independently of $(x^k_n)_{n=0}^N$, $k = 1, \dots, K$}
$(x^k_n)_{n=0}^N$, $k = K+1, \dots, K+K_L$, of $(X_n)_{n=0}^N$ and approximate 
$L$ with the Monte Carlo average 
\[
\hat{L} = \frac{1}{K_L} \sum_{k=K+1}^{K+K_L} g^k.
\]
Denote by $z_{\alpha/2}$ the $1-\alpha/2$ quantile of the standard normal distribution and consider 
the sample standard deviation 
\[
\hat{\sigma}_L = \sqrt{\frac{1}{K_L-1} \sum_{k=K+1}^{K+K_L} \brak{g^k -\hat{L}}^2}.
\]
Then one obtains from the central limit theorem that
\be \label{ciL}
\left[\hat{L} - z_{\alpha/2} \frac{\hat{\sigma}_L}{\sqrt{K_L}} \, , \, \infty \right)
\ee
is an asymptotically valid $1-\alpha/2$ confidence interval for $L$.

\subsection{Upper bound, point estimate and confidence intervals} 

Our derivation of an upper bound is based on the duality results of \cite{R02, HK04, DOS}. 
By \cite{R02, HK04}, the optimal value $V$ can be written as
\[
V = \mathbb{E} \edg{ \max_{0 \le n \le N} \brak{G_n - M_n}},
\]
where $(M_n)_{n=0}^N$ is the martingale part of the smallest $\F$-supermartingale dominating the payoff process $(G_n)_{n=0}^N$.
We approximate $(M_n)_{n = 0}^N$ with the $\mathbb{F}$-martingale $(M^{\Theta}_n)_{n=0}^N$ 
obtained from the stopping decisions implied by the trained continuation value functions $c^{\theta_n}$,
$n = 0 , \dots, N-1$, as in Section 3.2 of \cite{DOS}.
We know from Proposition 7 of \cite{DOS} that if $(\varepsilon_n)_{n=0}^N$ is a sequence 
of integrable random variables satisfying 
$\mathbb{E} \edg{\varepsilon_n \mid {\cal F}_n} = 0$ for all $n = 0, 1, \dots, N$, then 
\[
U = \mathbb{E} \edg{ \max_{0 \le n \le N} \brak{G_n - M^{\Theta}_n - \varepsilon_n}}
\]
is an upper bound for $V$. As in \cite{DOS}, we use nested simulation\footnote{The use of 
nested simulation ensures that $m^k_n$ are unbiased estimates of $M^{\Theta}_n$, which 
is crucial for the validity of the upper bound. In particular, we do not directly approximate $M^{\Theta}_n$
with the estimated continuation value functions $c^{\theta_n}$.} to generate realizations 
$m^k_n$ of $M^{\Theta}_n + \varepsilon_n$ along independent realizations $(x^k_n)_{n=0}^N$, 
$k = K + K_L + 1, \dots, K + K_L + K_U$, of $(X_n)_{n=0}^N$ sampled independently of 
$(x^k_n)_{n=0}^N$, $k = 1, \dots K$, and estimate $U$ as
\[
\hat{U} = \frac{1}{K_U} \sum_{k = K+ K_L + 1}^{K + K_L + K_U} \max_{0 \le n \le N} (g(n,x^k_n) - m^k_n).
\]
Our point estimate of $V$ is
\[
\hat{V} = \frac{\hat{L} + \hat{U}}{2}.
\]
The sample standard deviation of the estimator $\hat{U}$, given by
\[
\hat{\sigma}_U = \sqrt{\frac{1}{K_U-1} \sum_{k= K+K_L +1}^{K+K_L +K_U} \brak{\max_{0 \le n \le N} (g(n,x^k_n) - m^k_n)  -\hat{U}}^2},
\]
can be used together with the one-sided confidence interval \eqref{ciL} to construct the asymptotically 
valid two-sided $1- \alpha$ confidence interval
\be \label{ci}
\edg{\hat{L} - z_{\alpha/2} \frac{\hat{\sigma}_L}{\sqrt{K_L}} \,, \, \hat{U} 
+ z_{\alpha/2} \frac{\hat{\sigma}_U}{\sqrt{K_U}}}
\ee
for the true value $V$; see Section 3.3 of \cite{DOS}.

\section{Hedging}
\label{sec:hedging}

We now consider a savings account together with $e \in \N$ financial securities as hedging instruments.
We fix a positive integer $M$ and introduce a time grid $0 = u_1 < u_2 < \dots < u_{NM}$ such 
that $u_{nM} = t_n$ for all $n = 0,1, \dots, N$. We suppose that the information available at time
$u_m$ is described by ${\cal H}_m$, where $\IH = ({\cal H}_m)_{m = 0}^{MN}$ 
is a filtration satisfying ${\cal H}_{nM} = {\cal F}_n$ for all $n$. If any of the financial securities pay dividends,
they are immediately reinvested. We assume that the resulting discounted\footnote{Discounting is done
with respect to the savings account. Then, the discounted value of money  
invested in the savings account stays constant.} 
value processes are of the form $P_{u_m} = p_m(Y_m)$ for measurable functions $p_m \colon \R^d \to \R^e$ and 
an $\IH$-Markov process\footnote{That is, $Y_m$ is ${\cal H}_m$-measurable and 
$\mathbb{E}[f(Y_{m+1})\mid {\cal H}_m] = \mathbb{E}[f(Y_{m+1}) \mid Y_m]$ for all 
$m \le NM - 1$ and every measurable function $f \colon \R^d \to \R$ such that 
$f(Y_{m+1})$ is integrable.} $(Y_m)_{m =0}^{NM}$ such that $Y_{nM} = X_n$ for all 
$n = 0, \dots, N$. A hedging strategy consists of a sequence $h = (h_m)_{m=0}^{NM-1}$ of functions 
$h_m \colon \R^d \to \R^e$ specifying the time-$u_m$ holdings in $P^1_{u_m}, \dots, P^e_{u_m}$. 
As usual, money is dynamically deposited in or borrowed from the savings account to 
make the strategy self-financing. The resulting discounted gains at time $u_m$ are given by
\[
(h \cdot P)_{u_m} := \sum_{j = 0}^{m-1} h_j(Y_j) \cdot \brak{p_{j+1}(Y_{j+1})- p_j(Y_j)}
:= \sum_{j = 0}^{m-1} \sum_{i=1}^e h^i_j(Y_j) \brak{p^i_{j+1}(Y_{j+1} )- p^i_j(Y_j)}.
\]

\subsection{Hedging until the first possible exercise date}
\label{subsec:first}

For a typical Bermudan option, the time between two possible exercise dates $t_n - t_{n-1}$ 
might range between a week and several months. In case of an American option, we choose $t_n = n \Delta$ 
for a small amount of time $\Delta$ such as a day. We assume $\tau^{\Theta}$ does not stop at time $0$.
Otherwise, there is nothing to hedge. In a first step, we only compute the hedge until time $t_1$. If the option is still alive at time $t_1$,
the hedge can then be computed until time $t_2$ and so on. To construct a hedge from time $0$ 
to $t_1$, we approximate the time-$t_1$ value of the option with $V^{\theta_1}_{t_1} = v^{\theta_1}(X_1)$ 
for the function $v^{\theta_1}(x) = g(1,x) \vee c^{\theta_1}(x)$, where $c^{\theta_1} \colon \R^d \to \R$ 
is the time-$t_1$ continuation value function estimated in Section \ref{sec:st}.
Then we search for hedging positions $h_m$, $m = 0, 1, \dots, M-1$, that minimize the
mean squared error
\[
\E \edg{\brak{ \hat{V} + (h \cdot P)_{t_1} - V^{\theta_1}_{t_1} }^2}.
\]
To do that we approximate the functions $h_m$ with neural networks
$h^{\lambda} \colon \R^d \to \R^e$ of the form \eqref{nn} and try to find parameters
$\lambda_0, \dots, \lambda_{M-1}$ that minimize
\be \label{mse}
\sum_{k=1}^{K_H} \brak{ \hat{V} + \sum_{m = 0}^{M-1} h^{\lambda_m}(y^k_{m}) 
\cdot \brak{p_{m+1}(y^k_{m+1} )- p_{m}(y^k_{m})} - v^{\theta_1}(y^k_M)}^2
\ee
for independent realizations of $(y^k_m)_{m=0}^M$, $k=1, \dots, K_H$, of $(Y_m)_{m=0}^M$.
We train the networks $h^{\lambda_0}, \dots, h^{\lambda_{M-1}}$ together, again
using a stochastic gradient descent method. Instead of \eqref{mse}, one could also 
minimize a different deviation measure. But \eqref{mse} has the advantage that it yields 
hedging strategies with an average hedging error close to zero\footnote{see Table \ref{table:mchedge} 
and Figure \ref{fig} below}.

Once $\lambda_0, \dots, \lambda_{M-1}$ have been determined, we assess the 
quality of the hedge by simulating new\footnote{independent of $(y^k_m)_{m=0}^M$,
$k = 1, \dots, K_H$} independent realizations $(y^k_m)_{m=0}^M$, $k = K_H +1, \dots, K_H + K_E$, of $(Y_m)_{m=0}^M$
and calculating the average hedging error 
\be \label{ahe}
\frac{1}{K_E} \sum_{k = K_H+1}^{K_H+K_E} \brak{
\hat{V} + \sum_{m = 0}^{M-1} h^{\lambda_m}(y^k_m) 
\cdot \brak{p_{m+1}(y^k_{m+1} )- p_m(y^k_m)} - v^{\theta_1}(y^k_M)}
\ee
and the empirical hedging shortfall 
\be \label{ehs}
\frac{1}{K_E} \sum_{k = K_H+1}^{K_H+K_E} 
\brak{\hat{V} + \sum_{m = 0}^{M-1} h^{\lambda_m}(y^k_m) 
\cdot \brak{p_{m+1}(y^k_{m+1} )- p_m(y^k_m)} - v^{\theta_1}(y^k_M)}^-
\ee
over the time interval $[0,t_1]$.

\subsection{Hedging until the exercise time} 
\label{subsec:extime}

Alternatively, one can precompute the whole hedging strategy from time $0$ to $T$
and then use it until the option is exercised. In order to do that we introduce the functions 
\[
v^{\theta_n}(x) := g(n,x) \vee c^{\theta_n}(x), \quad
C^{\theta_n}(x) := 0 \vee c^{\theta_n}(x), \quad x \in \R^d,
\]
and hedge the difference $v^{\theta_n}(Y_{nM}) - C^{\theta_{n-1}}(Y_{(n-1)M})$ on each 
of the time intervals $[t_{n-1},t_n]$, $n = 1, \dots, N$, separately. 
$v^{\theta_n}$ describes the approximate value of the option at time $t_n$ if it has not been exercised before,
and the definition of $C^{\theta_n}$ takes into account that the true continuation values are non-negative 
due to the non-negativity of the payoff function $g$. The hedging strategy can be computed as in
Section \ref{subsec:first}, except that we now have to simulate
complete paths $(y^k_m)_{m=0}^{NM}$ of $(Y_m)_{m=0}^{NM}$, $k = 1, \dots, K_H$,
and then for all $n = 1, \dots, N$, find parameters 
$\lambda_{(n-1)M}, \dots, \lambda_{nM-1}$ which minimize
\[
\sum_{k=1}^{K_H} \brak{C^{\theta_{n-1}}(y^k_{(n-1)M}) + \sum_{m = (n-1)M}^{nM-1} 
h^{\lambda_m}(y^k_{m}) \cdot \brak{p_{m+1}(y^k_{m+1} )- p_{m}(y^k_{m})}
- v^{\theta_n}(y^k_{nM})}^2.
\]
Once the hedging strategy has been trained, we simulate independent samples$^9$
$(y^k_m)_{m = 0}^{NM}$, $k = K_H+1, \dots, K_H + K_E$, of $(Y_m)_{m=0}^{NM}$
and denote the realization of $\tau^{\Theta}$ along each sample path $(y^k_m)_{m=0}^{NM}$ 
by $\tau^k$. The corresponding average hedging error is given by 
\be \label{the}
\frac{1}{K_E} \sum_{k= K_H +1}^{K_H + K_E} 
\brak{\hat{V} + \sum_{m = 0}^{\tau^k M - 1}
h^{\lambda_m}(y^k_m) \cdot \brak{p_{m+1}(y^k_{m+1} )- p_m(y^k_m)}
- g(\tau^k, X_{\tau^k}) }
\ee
and the empirical hedging shortfall by
\be \label{ths}
\frac{1}{K_E} \sum_{k= K_H +1}^{K_H + K_E} 
\brak{\hat{V} + \sum_{m = 0}^{\tau^k M - 1}
h^{\lambda_m}(y^k_m) \cdot \brak{p_{m+1}(y^k_{m+1} )- p_m(y^k_m)}
- g(\tau^k, X_{\tau^k}) }^-.
\ee

\section{Example} 
\label{sec:ex}

In this section we study\footnote{The computations were performed
on a NVIDIA GeForce RTX 2080 Ti GPU. The underlying system was an AMD Ryzen 
9 3950X CPU with 64 GB DDR4 memory running Tensorflow 2.1 on Ubuntu 19.10.}
a Bermudan max-call option\footnote{Bermudan max-call options are a benchmark example
in the literature on numerical methods for high-dimensional American-style options; see, e.g.,
\cite{LS01, R02, G03, BG04, HK04, BC08, BS08, JO15, DOS, BCJW}.} 
on $d$ financial securities with risk-neutral price dynamics
\[
S^i_t = s^i_0 \exp\!\brak{[r-\delta_i - \sigma^2_i/2] t + \sigma_i W^i_t}, \quad i = 1, 2, \dots, d,
\]
for a risk-free interest rate $r \in \mathbb{R}$, initial values $s^i_0 \in (0,\infty)$, dividend yields
$\delta_i \in [0,\infty)$, volatilities $\sigma_i \in (0,\infty)$ and a $d$-dimensional Brownian motion $W$
with constant instantaneous correlations\footnote{That is,
$\mathbb{E}[(W^i_t-W^i_s)( W^j_t- W^i_s)] = \rho_{ij}(t-s)$ for all $i \neq j$ and $s < t$.} $\rho_{ij} \in \mathbb{R}$ 
between different components $W^i$ and $W^j$.
The option has time-$t$ payoff $\brak{\max_{1 \le i \le d} S^i_t - K}^+$ for a strike price $K \in [0,\infty)$
and can be exercised at one of finitely many times $0 = t_0 < t_1 < \dots < t_N = T$.
In addition, we suppose there is a savings account where money can be deposited and borrowed
at rate $r$.

For notational simplicity, we assume in the following that $t_n = nT/N$ for $n = 0,1,\dots, N$, and all assets have the 
same\footnote{Simulation based methods work for any price dynamics that can efficiently be simulated.
Prices of max-call options on underlying assets with different price dynamics were calculated in \cite{BC08} and \cite{DOS}.} 
characteristics; that is, $s^i_0 = s_0$, $\delta_i = \delta$ and $\sigma_i = \sigma$ 
for all $i = 1, \dots, d$.

\subsection{Pricing results} 
\label{subsec:pr}

Let us denote $X_n = S_{t_n}$, $n = 0,1,\dots, N$. Then the price of the option is given by
\[
\sup_{\tau} \mathbb{E}\!\left[ e^{ - r \frac{\tau T}{N} } \left( 
\max_{ 1 \le i \le d } X^i_{ \tau } - K \right)^{ \! + } \right] ,
\]
where the supremum is over all stopping times $\tau \colon \Omega \to \{0,1,\dots, N\}$
with respect to the filtration generated by $(X_n)_{n=0}^N$. The option payoff does not carry 
any information not already contained in $X_n$. But the training of the continuation values worked more 
efficiently when we used it as an additional feature. So instead of $X_n$ 
we simulated the extended state process $\hat{X}_n = (X^1_n, \dots, X^d_n, X^{d+1}_n)$ for 
\[
X^{d+1}_n = e^{ - r \frac{n T}{N}} \left(\max_{ 1 \le i \le d } X^i_{ n } - K \right)^{ \! + }
\]
to train the continuation value functions $c^{\theta_n}$, $n =1, \dots, N-1$.
The network $c^{\theta} \colon \R^{d+1} \to \R$ was chosen of the form \eqref{nn} with depth 
$I=3$ (two hidden layers), $d+ 50$ nodes in each hidden layer and activation function $\varphi = \tanh$.
For training we used stochastic gradient descent with mini-batches of size 8,192 and batch normalization \cite{IS15}.
At time $N-1$ we used Xavier \cite{GB10} initialization and performed 6,000 Adam \cite{KiB15} 
updating steps\footnote{The hyperparamters $\beta_1, \beta_2, \varepsilon$ were chosen as in \cite{KiB15}. 
The stepsize $\alpha$ was specified as $10^{-1}$, $10^{-2}$, $10^{-3}$ and $10^{-4}$
according to a deterministic schedule.}.
For $n \le N-2$, we started the gradient descent from the trained network parameters $\theta_{n+1}$
and made 3,500 Adam \cite{KiB15} updating steps$^{16}$. To calculate $\hat{L}$ we simulated
$K_L =$ 4,096,000 paths of $(X_n)_{n =0}^N$. For $\hat{U}$ we generated $K_U =$ 2,048 outer and 
2,048 $\times$ 2,048 inner simulations.

Our results for $\hat{L}$, $\hat{U}$, $\hat{V}$ and 95\% confidence intervals 
for different specifications of the model parameters are reported in Table \ref{table:mcprices}.
To achieve a pricing accuracy comparable to the more direct methods of 
\cite{DOS} and \cite {BCJW}, the networks used in the construction of the candidate optimal 
stopping strategy had to be trained for a longer time. But in exchange, the approach yields 
approximate continuation values that can be used to break down the hedging problem 
into a series of smaller problems.

\begin{table}
\centering
\begin{small}
\begin{tabular}{c c c c c c c c c c} 
 \hline \\[-3mm]
  $d$ & $s_0$ & $\hat{L}$ & $t_L$ & $\hat{U}$ & $t_U$ & Point Est.\ & $95\%$ CI & DOS $95\%$ CI\\[1mm]
 \hline \\[-3mm]
 $5$ & $90$ & $16.644$ & $132$ & $16.648$ & $8$ & $16.646$ & $[16.628, 16.664]$ & $[16.633, 16.648]$\\
 $5$ & $100$ & $26.156$ & $134$ & $26.152$ & $8$ & $26.154$ & $[26.138, 26.171]$ & $[26.138, 26.174]$\\
 $5$ & $110$ & $36.780$ & $133$ & $36.796$ & $8$ & $36.788$ & $[36.758, 36.818]$ & $[36.745, 36.789]$\\[1mm]
 $10$ & $90$ & $26.277$ & $136$ & $26.283$ & $8$ & $26.280$ & $[26.259, 26.302]$ & $[26.189, 26.289]$\\
 $10$ & $100$ & $38.355$ & $136$ & $38.378$ & $7$ & $38.367$ & $[38.335, 38.399]$ & $[38.300, 38.367]$\\
 $10$ & $110$ & $50.869$ & $135$ & $50.932$ & $8$ & $50.900$ & $[50.846, 50.957]$ & $[50.834, 50.937]$\\[1mm]
 \hline
\end{tabular}
\end{small}
\caption{\label{table:mcprices}Price estimates for max-call options on $5$ and 10 symmetric assets for 
parameter values of $r = 5\%$, $\delta = 10\%$, $\sigma = 20\%$, $\rho = 0$, $K=100$, $T=3$, $N=9$.
$t_L$ is the number of seconds it took to train $\tau^{\Theta}$ and compute $\hat{L}$.
$t_U$ is the computation time for $\hat{U}$ in seconds. 95\% CI is the 95\% confidence interval \eqref{ci}. 
The last column lists the 95\% confidence intervals computed in \cite{DOS}.}
\end{table}

\subsection{Hedging results}

Suppose the hedging portfolio can be rebalanced at the times
$u_m = mT/(NM)$, $m = 0,1, \dots, NM$, for a positive integer $M$. We assume dividends paid by
shares of $S^i$ held in the hedging portfolio are continuously reinvested in $S^i$.
This results in the adjusted discounted security prices
\[
P^i_{u_m} = s_0 \exp\!\brak{\sigma W^i_{u_m} - \sigma^2 u_m/2}, \quad
m = 0,1, \dots, NM.
\]
We set $Y^i_m = P^i_{u_m}$. To learn the hedging strategy, we trained neural networks 
$h^{\lambda_m} \colon \R^d \to \R^d$, $m = 0, \dots, NM -1$, of the form \eqref{nn} with depth $I = 3$
(two hidden layers), $d+50$ nodes in each hidden layer and activation function $\varphi =\tanh$. 
As in Section \ref{subsec:pr}, we used stochastic gradient descent with mini-batches of size 
8,192 and batch normalization \cite{IS15}. For $m = 0, \dots, M-1$, we initialized the networks
according to Xavier \cite{GB10} and performed 10,000 Adam \cite{KiB15} 
updating steps$^{16}$, whereas for $m \ge M$, we started the gradient trajectories from 
the trained network parameters $\lambda_{m -M}$ and made 3,000 Adam \cite{KiB15} updating steps$^{16}$.

Table \ref{table:mchedge} reports the average hedging errors \eqref{ahe} and \eqref{the} together with the
empirical hedging shortfalls \eqref{ehs} and \eqref{ths} for different numbers $M$ of rebalancing times between two 
consecutive exercise dates $t_{n-1}$ and $t_n$. They were computed using $K_E =$ 4,096,000 simulations of 
$(Y_m)_{m = 0}^{NM}$.

\begin{table}[p]
\centering
\begin{small}
\begin{tabular}{c c c c c c c c c c c} 
 \hline \\[-3mm]
  $d$ & $s_0$ & $M$ & IHE & IHS & IHS/$\hat{V}$ & T1 & HE & HS & HS/$\hat{V}$ & T2 \\[1mm]
 \hline \\[-3mm]
$5$ & $90$ & $12$ & $0.007$ & $0.190$ & $1.1\%$ & $102$ & $-0.001$ & $0.676$ & $4.1\%$ & $379$ \\
$5$ & $90$ & $24$ & $0.007$ & $0.139$ & $0.8\%$ & $129$ & $-0.002$ & $0.492$ & $3.0\%$ & $473$ \\
$5$ & $90$ & $48$ & $0.007$ & $0.104$ & $0.6\%$ & $234$ & $-0.001$ & $0.367$ & $2.2\%$ & $839$ \\
$5$ & $90$ & $96$ & $0.007$ & $0.081$ & $0.5\%$ & $436$ & $-0.001$ & $0.294$ & $1.8\%$ & $1{,}546$ \\[1mm]
$5$ & $100$ & $12$ & $0.013$ & $0.228$ & $1.4\%$ & $102$ & $0.006$ & $0.785$ & $4.7\%$ & $407$ \\
$5$ & $100$ & $24$ & $0.013$ & $0.163$ & $1.0\%$ & $131$ & $0.006$ & $0.569$ & $3.4\%$ & $512$ \\
$5$ & $100$ & $48$ & $0.013$ & $0.118$ & $0.7\%$ & $252$ & $0.007$ & $0.423$ & $2.5\%$ & $931$ \\
$5$ & $100$ & $96$ & $0.013$ & $0.089$ & $0.5\%$ & $470$ & $0.006$ & $0.335$ & $2.0\%$ & $1{,}668$ \\[1mm]
$5$ & $110$ & $12$ & $0.002$ & $0.268$ & $1.6\%$ & $102$ & $-0.012$ & $0.881$ & $5.3\%$ & $380$ \\
$5$ & $110$ & $24$ & $0.002$ & $0.192$ & $1.2\%$ & $130$ & $-0.012$ & $0.638$ & $3.8\%$ & $511$ \\
$5$ & $110$ & $48$ & $0.002$ & $0.139$ & $0.8\%$ & $262$ & $-0.013$ & $0.474$ & $2.9\%$ & $950$ \\
$5$ & $110$ & $96$ & $0.002$ & $0.105$ & $0.6\%$ & $471$ & $-0.010$ & $0.374$ & $2.3\%$ & $1{,}673$ \\[1mm]
$10$ & $90$ & $12$ & $-0.015$ & $0.192$ & $0.7\%$ & $111$ & $-0.010$ & $0.902$ & $3.4\%$ & $414$ \\
$10$ & $90$ & $24$ & $-0.014$ & $0.147$ & $0.6\%$ & $145$ & $-0.011$ & $0.704$ & $2.7\%$ & $534$ \\
$10$ & $90$ & $48$ & $-0.015$ & $0.136$ & $0.5\%$ & $269$ & $-0.011$ & $0.611$ & $2.3\%$ & $958$ \\
$10$ & $90$ & $96$ & $-0.015$ & $0.121$ & $0.5\%$ & $506$ & $-0.012$ & $0.551$ & $2.1\%$ & $1{,}792$ \\[1mm]
$10$ & $100$ & $12$ & $0.008$ & $0.230$ & $0.9\%$ & $111$ & $0.015$ & $1.025$ & $3.9\%$ & $414$ \\
$10$ & $100$ & $24$ & $0.008$ & $0.176$ & $0.7\%$ & $152$ & $0.014$ & $0.797$ & $3.0\%$ & $531$ \\
$10$ & $100$ & $48$ & $0.008$ & $0.150$ & $0.6\%$ & $271$ & $0.016$ & $0.682$ & $2.6\%$ & $978$ \\
$10$ & $100$ & $96$ & $0.008$ & $0.132$ & $0.5\%$ & $512$ & $0.014$ & $0.672$ & $2.6\%$ & $1{,}803$ \\[1mm]
$10$ & $110$ & $12$ & $-0.029$ & $0.249$ & $1.0\%$ & $112$ & $-0.026$ & $1.146$ & $4.4\%$ & $410$ \\
$10$ & $110$ & $24$ & $-0.029$ & $0.189$ & $0.7\%$ & $146$ & $-0.027$ & $0.908$ & $3.5\%$ & $530$ \\
$10$ & $110$ & $48$ & $-0.029$ & $0.160$ & $0.6\%$ & $269$ & $-0.026$ & $0.782$ & $3.0\%$ & $965$ \\
$10$ & $110$ & $96$ & $-0.029$ & $0.151$ & $0.6\%$ & $507$ & $-0.024$ & $0.666$ & $2.5\%$ & $1{,}777$ \\[1mm]
 \hline
\end{tabular}
\end{small}
\caption{\label{table:mchedge}
Average hedging errors and empirical hedging shortfalls for $5$ and 10 underlying assets and different numbers $M$ of 
rehedging times between consecutive exercise times $t_{n-1}$ and $t_n$. The values of
the parameters $r$, $\delta$, $\sigma$, $\rho$, $K$, $T$ and $N$ were chosen as in Table \ref{table:mcprices}.
IHE is the intermediate average hedging error \eqref{ahe}, IHS the intermediate hedging shortfall \eqref{ehs},
HE the total average hedging error \eqref{the} and HS the total hedging shortfall \eqref{ths}. 
$\hat{V}$ is our price estimate from Table \ref{table:mcprices}.
T1 is the computation time in seconds for training the hedging strategy from time $0$ to $t_1 = T/N$.
T2 is the number of seconds it took to train the complete hedging strategy from time 0 to $T$.}
\end{table}

\begin{figure} 
\centering
\includegraphics[scale=0.95, trim=0cm 1cm 0cm 0cm]{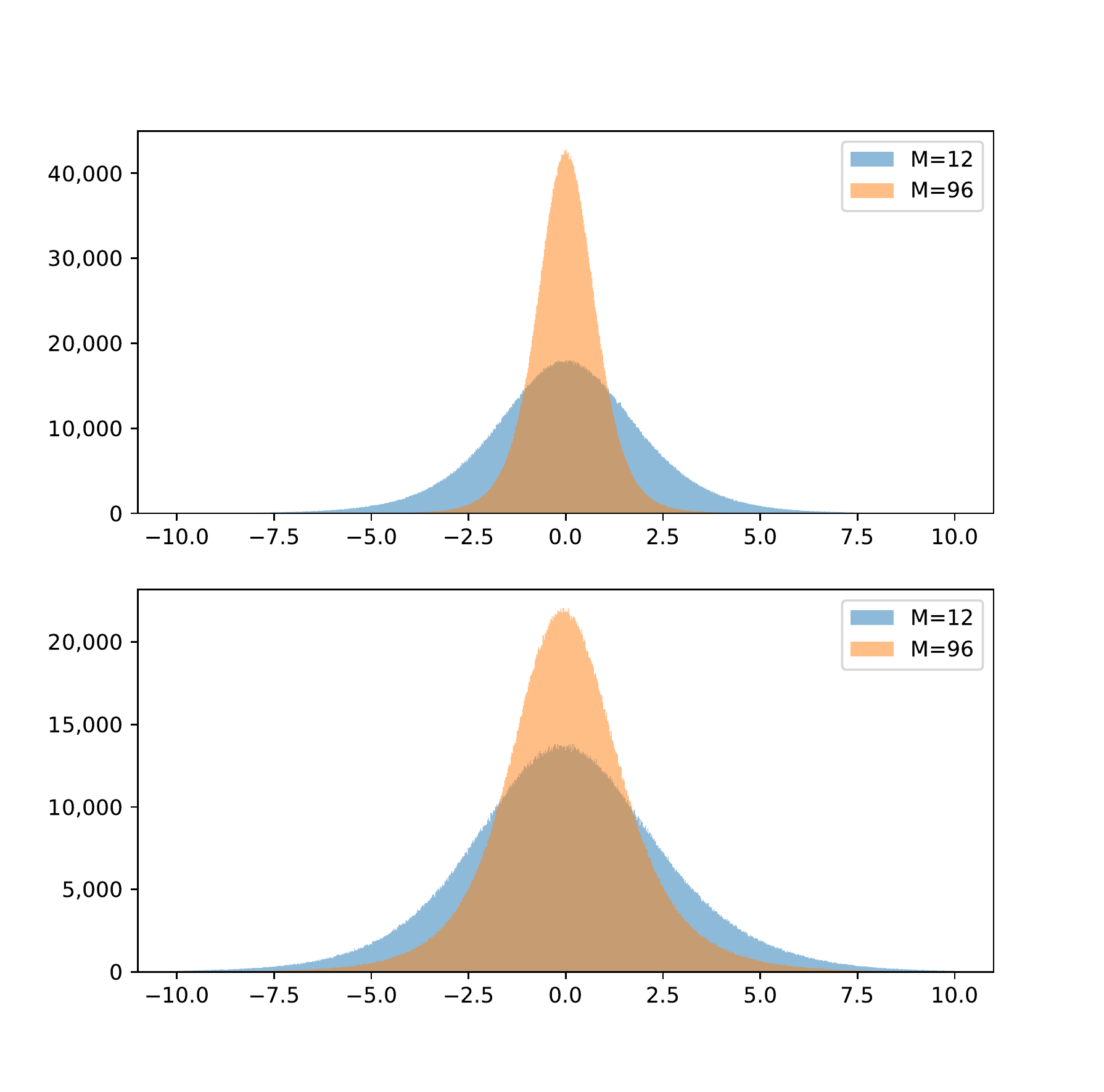}
\caption{\label{fig}Total hedging errors for $s_0=100$, $M \in \{12, 96\}$, $d = 5$ (top) and $d = 10$ (bottom)
along 4,096,000 sample paths of $(Y_m)_{m=0}^{NM}$. The parameters $r$, $\delta$, $\sigma$, 
$\rho$, $K$, $T$ and $N$ were specified as in Tables \ref{table:mcprices} and \ref{table:mchedge}.} 
\end{figure}

Figure \ref{fig} shows histograms of the total hedging errors 
\[
\hat{V} + \sum_{m = 0}^{\tau^k M - 1}
h^{\lambda_m}(y^k_m) \cdot \brak{p_{m+1}(y^k_{m+1} )- p_m(y^k_m)}
- g(\tau^k, X_{\tau^k}), \quad k = K_H + 1, \dots, K_E,
\]
for $d \in \crl{5,10}$ and $M \in \crl{12, 96}$.

\section{Conclusion}
\label{sec:conclusion}

In this article we used deep learning to price and hedge American-style options.
In a first step our method employs a neural network version of the Longstaff--Schwartz algorithm 
to estimate continuation values and derive a candidate optimal stopping rule. The learned 
stopping rule immediately yields a low-biased estimate of the price. In addition, it can be used to 
construct an approximate solution of the dual martingale problem of \cite{R02, HK04}.
This gives a high-biased estimate and confidence intervals for the price. To achieve 
the same pricing accuracy as the more direct approaches of \cite{DOS} and \cite{BCJW},
we had to train the neural network approximations of the continuation values for a longer time.
But computing approximate continuation values has the advantage that they can be used 
to break the hedging problem into a sequence of subproblems that compute the hedge 
only from one possible exercise date to the next.

\end{document}